\def\beq{\begin{equation}}
\def\eeq{\end{equation}}
\def\beqa{\begin{eqnarray}}
\def\eeqa{\end{eqnarray}}
\def\eq#1{Eq.~(\ref{#1})}

\def\slash#1{#1 \hskip-0.45em /}

\documentclass{PoS}

\title{Next-to-leading power threshold logarithms: a status report}

\ShortTitle{NLP Logarithms}

\author{\speaker{Lorenzo Magnea}\\
        Universit\`a di Torino, and INFN, Sezione di Torino\\
        E-mail: \email{lorenzo.magnea@unito.it}}

\author{Domenico Bonocore\\
        Nikhef,  Amsterdam\\
        E-mail: \email{d.bonocore@nikhef.nl}}

\author{Eric Laenen\\
        Nikhef, Amsterdam, and ITFA, University of Amsterdam, and ITF, Utrecht University\\
        E-mail: \email{t45@nikhef.nl}}

\author{Leonardo Vernazza\\
        Higgs Centre for Theoretical Physics, The University of Edinburgh\\
        E-mail: \email{lvernazz@staffmail.ed.ac.uk}}

\author{Chris D. White and Stacey Melville\\
        School of Physics and Astronomy, University of Glasgow\\
        E-mail: \email{Christopher.White@glasgow.ac.uk, s.melville.1@research.gla.ac.uk}}

\abstract{There is ample evidence, dating as far back as Low's theorem, that the universality
of soft emissions extends beyond leading power in the soft energy. This universality can, in 
principle, be exploited to generalise the formalism of threshold resummations beyond leading
power in the threshold variable. In the past years, several phenomenological approaches
have been partially successful in performing such a resummation. Here, we briefly review 
some recent developments which pave the way to a solution of this problem, at least
for electroweak annihilation processes.}

\FullConference{12th International Symposium on Radiative Corrections (Radcor 2015) 
and LoopFest XIV (Radiative Corrections for the LHC and Future Colliders)\\
15-19 June, 2015\\UCLA Department of Physics and Astronomy Los Angeles, USA}

\begin{document}

\section{Introduction}
\label{intro}

Standard Model cross sections of interest for LHC often involve many different physical 
scales, including for example Mandelstam invariants, heavy particle masses and jet 
masses. Since the Standard Model is a renormalisable gauge theory, perturbative 
expressions for these cross section always involve powers of logarithms of ratios of 
these scales. When some of the scales are disparate, the logarithms are large, and they 
often spoil the reliability of perturbation theory in phenomenologically relevant kinematic regions.
In many cases, these logarithms are associated with underlying singularities of scattering 
amplitudes, and in particular they often appear as finite remainders after the cancellation
of divergences arising at the edges of phase space. As a consequence, such logarithms 
have a universal nature, in the sense that they do not depend on the details of the hard 
scattering process at hand. They can then be computed once and for all, and often 
formally summed up to all orders in perturbation theory, yielding improved predictions
for physical observables which can be applied to more extreme configurations. Well-known
examples of this `resummation' technology are given by the renormalisation group,
by perturbative Reggeization, and by Sudakov resummation.

In this contribution, we will be concerned with a specific class of these logarithms, 
arising when a partonic cross section is evaluated in the vicinity of a physical threshold
for the production of a selected final state.  A slightly unconventional way to define these
`threshold logarithms' is the following: they are those that arise in distributions which, 
at Born level, are localised at the threshold (in other words, the Born distribution is 
a delta function). This definition thus includes $p_t$ logarithms arising, say, in
the Drell-Yan or Higgs $p_t$ distributions, as well as conventional Sudakov logarithms 
in the inclusive cross sections for electroweak annihilation processes, DIS, and 
event shapes in electron-positron annihilation.

In all these cases, one can define a "threshold variable" $\xi$, such that the Born 
cross section is proportional to $\delta(\xi)$. Loop corrections generically take the 
form
\beq
  \frac{d \sigma}{d \xi} \, = \, \sum_{n = 0}^{\infty} \left( \frac{\alpha_s}{\pi} \right)^n \, 
  \sum_{m = 0}^{2 n - 1} \left[ c_{n m}^{(-1)}
  \left( \frac{\log^m \xi}{\xi} \right)_+ + \, c_{nm}^{(\delta)} \, \delta(\xi) \, + \, 
  c_{nm}^{(0)} \, \log^m \xi \, + \, \ldots \right] \, ,
\label{thresholddef}
\eeq
where the `plus distribution' notation is used here generically to indicate that one must 
include virtual corrections in order for the first set of terms to be integrable. The 
logarithmic terms with coefficients given by $c_{n m}^{(-1)}$ are the conventional
Sudakov logarithms, closely connected to infrared and collinear divergences of the 
relevant amplitudes. Their resummation has been well understood for many years,
and is routinely applied, to high logarithmic accuracy, for a wide range of observables.
In the present context, we refer to these terms as {\it leading-power} (LP) threshold
logarithms. The second set of terms, with coefficients given by $c_{nm}^{(\delta)}$,
arises from finite virtual corrections and from remainders of phase space integrations
after the cancellation of IR and collinear poles. Interestingly, as we will briefly review 
below, for cross sections that are purely electroweak at tree level these terms can
also be studied to all orders in perturbation theory, albeit with a lesser degree of 
control as compared with LP logarithms. Finally, the last set of terms in \eq{thresholddef},
with coefficients given by $c_{nm}^{(0)}$, is the main subject of this contribution: 
these terms are integrable, but they can still give significant contributions to the cross 
section, order by order in perturbation theory, when $\xi$ is small. We refer to these
terms as {\it next-to-leading-power} (NLP) threshold logarithms. Over the years,
an increasing body of evidence has accumulated, suggesting that NLP logarithms 
can be organised to all orders in perturbation theory, similarly to what happens
at LP. Our goal here is to briefly review this body of evidence, and then summarise
some very recent results which were presented in detail in Refs.~\cite{Bonocore:2014wua,
Bonocore:2015esa}\footnote{See also~\cite{Bonocore:2015anl}.}.

\section{Gathering evidence}
\label{evide}

Beyond LP threshold logarithms, the first interesting contributions to the cross 
section are those which are localised at the threshold, specified by the coefficients
$c_{nm}^{(\delta)}$ in \eq{thresholddef}. In dimensional regularisation, and for 
processes which are electroweak at tree level, {\it all} these terms are naturally 
organised in exponential form, as a consequence of the evolution equations
obeyed by the various factors composing the partonic cross section. The first 
observation in this direction dates back to~\cite{Parisi:1979xd}, and has been successively
refined, extended and revisited in~\cite{Sterman:1986aj,Eynck:2003fn,Ahrens:2008qu}.
Following the reasoning of~\cite{Eynck:2003fn}, and using the Drell-Yan process
as an example, one may simply note that the partonic cross section for 
quark-initiated Drell-Yan near threshold obeys the (Mellin space) factorisation 
theorem~\cite{Sterman:1986aj}
\beq
  \omega (N, \epsilon) \, = \, \left| \Gamma \left( Q^2, \epsilon \right) \right|^2 \, 
  \left[ \psi_R (N, \epsilon) \right]^2 \, U_R (N, \epsilon) + {\cal O} \left( \frac{1}{N} \right) \, ,
\label{factoG}
\eeq
where $\Gamma$ is the quark form factor, and $\psi_R$ and $U_R$ are responsible
respectively for collinear and soft {\it real} radiation into the final state. In \eq{factoG}
real and virtual correction are treated separately: this is possible only because each 
factor obeys evolution equations which can be solved in exponential form, with trivial
boundary conditions in dimensional regularisation. Infrared divergences cancel between 
the real emission function $U_R$ and the virtual form factor, while collinear divergences
remain in the parton distribution $\psi_R$, in factorised form. To construct the finite
partonic Drell-Yan cross section in Mellin space, in the ${\overline{MS}}$ scheme, it 
is now sufficient to divide \eq{factoG} by the square of the ${\overline{MS}}$
parton distribution $\phi_{\overline{MS}}$, which can also be written as the product 
of a virtual factor times a real emission factor. One is led to the expression
\beq
  \widehat{\omega}_{\overline{MS}} (N) \, = \, \left( 
  \frac{\left| \Gamma \left( Q^2, \epsilon \right) \right|^2}{\left[ \phi_V(\epsilon) \right]^2} \right) \, 
  \left( \frac{\left[ \psi_R (N, \epsilon) \right]^2 \, U_R (N, \epsilon)}{\left[ \phi_R (N, \epsilon) 
  \right]^2} \right) + {\cal O} \left( \frac{1}{N} \right) \, .
\label{finfacto}
\eeq
Since each factor in \eq{finfacto} exponentiates, and the factorisation is accurate 
up to NLP corrections, it follows that constant terms in Mellin space 
(corresponding to localised terms in momentum space) are naturally defined 
in the exponent. Clearly, the predictive power of this statement is limited: when 
exponentiating logarithms, a finite-order calculation makes an exact prediction 
for a set of infinite towers of logarithmic corrections to all orders in perturbation 
theory; constants, on the other hand, cannot be categorised by parametric 
enhancements: therefore, at order $n$, they receive contributions both from 
the exponentiation of lower orders and from terms genuinely arising at order 
$n$. This not withstanding, it is certainly legitimate to use exponentiation at 
least as a tool to estimate the size of higher-order localised corrections.

At NLP level, the first historical bit of evidence for the universality of logarithmic 
corrections is Low's theorem~\cite{Low:1958sn}, to be reviewed in the next 
section. It is however non-trivial to make use of Low's theorem to construct 
a resummation formalism. A more immediately applicable proposal was made 
in~\cite{Dokshitzer:2005bf}, building upon an empirical observation arising 
from the three-loop calculation of Ref.~\cite{Moch:2004pa}. The central physical 
input of~\cite{Dokshitzer:2005bf} is {\it reciprocity}, the idea that the evolution 
kernels for parton splitting and fragmentation should be simply related by analytic 
continuation. This idea can be realised by making use of a modified evolution 
equation, the DMS equation, which can be written as
\beq
  \mu^2 \frac{\partial}{\partial \mu^2} \, \psi (x, \mu^2) \, = \, 
  \int_x^1 \frac{dz}{z} \,  \psi \left( \frac{x}{z}, z^\sigma \mu^2 \right) 
  {\cal P} \left( z, \alpha_s \left( \frac{\mu^2}{z} \right) \right) \, ,
\label{dms}
\eeq
where $\sigma = +1$ for space-like evolution of parton densities, and $\sigma 
= - 1$ for time-like evolution of fragmentation functions. Ref.~\cite{Dokshitzer:2005bf}
argues that, in a renormalisation scheme where the coupling is defined to equal
the light-like cusp anomalous dimension, the universal kernel ${\cal P}$ has
the remarkable property of having no corrections at NLP in $(1-z)$. In other
words, one may write
\beq
  {\cal P} \left( z, \alpha_s \right) \, = \, \frac{A(\alpha_s)}{(1 - z)_+} +
  B_\delta (\alpha_s) \delta (1 - z) + {\cal O} (1 - z) \, ,
\label{noNLP}
\eeq
a relation which is verified up to three loops in QCD. Clearly, \eq{dms} cannot
be solved as easily as the ordinary DGLAP equation, since it is not diagonalised 
by a Mellin transform. It is however possible to solve it recursively, order by 
order in perturbation theory. Proceeding in this way, and using \eq{noNLP},
one can map DMS evolution into ordinary DGLAP evolution, with a modified
kernel such that higher-order coefficients of NLP contributions are determined
by lower-order coefficients of the anomalous dimensions $A (\alpha_s)$ and 
$B_\delta (\alpha_s)$,  explaining and generalising the observation of
Ref.~\cite{Moch:2004pa}.

These insights can easily be combined to construct an improved threshold 
resummation formula, including in the perturbative exponent a subset of NLP 
logarithms, as well as contributions localised at threshold. This was done
in Ref.~\cite{Laenen:2008ux}, where the following expression was proposed 
for the Drell-Yan cross section
\beqa
  \ln \Big[ \widehat{\omega} (N) \Big] & = & 
  {\cal F}_{\rm DY} \left( \alpha_s (Q^2) \right) + 
  \int_0^1 \, dz \, z^{N - 1} \, \Bigg\{ \frac{1}{1 - z} \, 
  D \left[ \alpha_s \left( \frac{(1 - z)^2 Q^2}{z} \right) \right] \nonumber \\ & &
  + \,\,  2 \,\int_{Q^2}^{(1 - z)^2 Q^2/z} \, 
  \frac{d q^2}{q^2} \, P_s \Big[ z, \alpha_s (q^2) \Big] \Bigg\}_+ \, .
\label{newresDY}
\eeqa
Eq.~(\ref{newresDY}) improves upon standard LP threshold resummation in three ways:
first, contributions localized at threshold are included in the exponent, collected in the
function ${\cal F}_{DY} (\alpha_s)$; second, phase space limits for soft radiation are
evaluated to higher accuracy in $(1-z)$, both in the argument of the coupling in the soft 
function $D(\alpha_s)$ and in the limit of integration\footnote{A similar improvement of
resummation was proposed in \cite{Ball:2013bra} for Higgs production in the gluon fusion 
channel, where it was coupled with information coming from the high-energy ($N \to 1$) 
limit.}; third, in the leading term the cusp anomalous dimension is replaced by the 
DMS-improved splitting function $P_s$. Explicit comparison of \eq{newresDY} with 
finite order results at two and three loops shows that leading and next-to-leading NLP 
logarithms at higher orders are predicted with remarkable accuracy, based on lower 
order results: for example, leading NLP logarithms at two loops can be generated by 
the simple substitution $2/(1-z) \to 2 z/(1-z)$ in the cusp term, as was noticed already 
in Ref.~\cite{Kramer:1996iq}. Further subleading NLP logarithms, however, are predicted 
with decreasing accuracy, and it is clear that a more systematic approach is necessary 
in order to achieve a reliable and complete resummation. Steps towards such an 
approach are described in the next two sections.

\section{Towards systematics}
\label{syste}

Over the past several years, a number of approaches have been developed 
to improve our understanding of NLP logarithms in hadronic cross sections. 
The literature is vast and cannot be reviewed here, but, to mention the most 
recent delopments, the physical kernel method developed by Moch and 
Vogt~\cite{Moch:2009hr} has been recently applied to the Higgs production cross section 
in~\cite{deFlorian:2014vta}, and Soft-Collinear Effective Theory has been applied 
to this problem in~\cite{Larkoski:2014bxa,Kolodrubetz:2016uim}. In a massless 
theory, such as perturbative QCD in most applications, the challenge of constructing
a general formalism is dual: first, one must study how the formalism of soft gluon
factorisation and exponentiation generalises beyond leading power; then one
must include in the picture (next-to-) collinear configurations, which may (and do)
interfere with the soft expansion.

The task of extending the well-known soft factorisation and exponentiation theorems
beyond leading power was first tackled in~\cite{Laenen:2008gt}, using a path integral 
formalism. Neglecting collinear problems, and using techniques similar to world-line
methods, it is easy to see how the eikonal approximation arises in this context. The
replica trick often used in statistical field theory then leads to exponentiation at 
eikonal level. These methods can be extended to next-to-leading power in the soft 
energy, sometimes called Next-to-Eikonal (NE): the result is that a large set 
of contributions to scattering amplitudes factorise and exponentiate, and the
exponent of the next-to-soft factor can be directly computed in terms of NE 
Feynman rules. A non-factorizable remainder survives, which in this context is 
partly associated with translations of the relevant Wilson lines. Using Ward identities,
this setup can be transparently mapped back to Low's theorem in the simple
case of photon emission. The same conclusion can be confirmed from a purely
diagrammatic point of view, which was pursued in Ref.~\cite{Laenen:2010uz}.
With this method, the problem is straightforward in principle, but requires a very 
intricate combinatorial analysis. The starting point is the expansion of the propagator 
of the particle carrying the hard momentum $p$, and emitting the soft gluon, in 
powers of the soft gluon momentum $k$. For a massless spin one-half emitter, 
one writes
\beq
  \frac{\slash{p} + \slash{k}}{2 p \cdot k + k^2} \gamma^\mu u(p) \, = \, 
  \left[ \frac{p^\mu}{p \cdot k} - k^2 \frac{p^\mu}{2 \left(p \cdot k\right)^2}
  + \frac{\slash{k} \gamma^\mu}{2 p \cdot k} \right] u(p) \, + \, {\cal O} (k) \, ,
\label{neik}
\eeq
where one recognises the eikonal vertex at leading power in $k$, followed 
by a spin-independent next-to-soft term, and finally by a spin-dependent
contribution. Indeed, it is well-known since Low's days~\cite{Burnett:1967km} 
that the universality (and in particular the spin-independence) of soft emissions 
breaks down at NLP in the soft energy. At the level of matrix elements, the 
results of Refs.~\cite{Laenen:2008gt,Laenen:2010uz} can be summarised 
as follows.

In the eikonal approximation, it is well known that soft emissions factorise from matrix 
elements, and the resulting soft function can be written as a correlator of 
Wilson lines (see, for example,~\cite{Dixon:2008gr,Gardi:2009zv}). Furthermore, 
it is known that the soft function exponentiates, and the exponent can be directly 
computed in terms of a subset of the original Feynman diagrams~\cite{Gardi:2010rn,
Mitov:2010rp}. For a correlator of $n$ Wilson lines, one writes
\beq
S_n \equiv \langle 0 | \Phi_1 \otimes  \ldots \otimes \Phi_n | 0 \rangle \, = \, 
\exp \left( w_n \right) \, .
\eeq
If one then expresses each diagram $D$ contributing to $S_n$ as the product of a 
color factor $C(D)$ and a kinematic factor ${\cal F} (D)$, one finds that $w_n$
can be written, order by order in the coupling, as a sum over a subset of the
diagrams $D$, organized in structures called {\it webs}. Each web is a set
of diagrams differing by the order of gluon attachments on the Wilson lines,
and computed with modified color factors, according to
\beq
W \, = \, \sum_{D \in W} \widetilde{C} (D) {\cal F} (D) \, = \, 
\sum_{D, D' \in W} C (D') \, R \left(D', D\right) {\cal F} (D) \, ,
\label{web}
\eeq
where $R(D',D)$, the {\it web mixing matrix}, is a matrix of constant combinatorial 
coefficients which can be computed recursively~\cite{Gardi:2011wa}. In this language, 
Refs.~\cite{Laenen:2008gt,Laenen:2010uz} show that matrix elements retain a similar 
structure at NLP. One may formally write
\beq
  {\cal M} \, =  \, {\cal M}_0 \, \exp \left[ \sum_{D_E} \widetilde{C} \left( D_E \right)
  {\cal F} \left( D_E \right) + \sum_{D_{NE}} \widetilde{C} \left( D_{NE} \right)
  {\cal F} \left( D_{NE} \right) \right] \, + \, {\cal M}_R \, + \, {\cal O} \left( NNE \right) \, ,
\label{neikexp}
\eeq
where $D_E$ are the diagrams (forming webs) that would appear in the eikonal 
approximation, while $D_{NE}$ are a set of diagrams constructed with new, 
next-to-eikonal Feynman rules. As might have been expected, factorisation and 
exponentiation are incomplete at NLP, and a non-factorisable remainder ${\cal M}_R$
survives, which can be computed order by order using Low's theorem.

Eq.~(\ref{neikexp}), with the appropriate NE Feynman rules, was tested in 
Ref.~\cite{Laenen:2010uz} by reproducing (at NLP) the two loop results for the 
double real emission contribution to the Drell-Yan cross section. It became clear, 
however, that the same technique fails when attempting to compute real-virtual
corrections. The reason can be traced to a failure of Low's theorem for massless
particles, which we discuss in the next section.

\section{Factorization at NLP level}
\label{facto}

The original version of Low's theorem~\cite{Low:1958sn}, later generalised to 
particles with spin by Burnett and Kroll~\cite{Burnett:1967km}, was derived for
massive particles, and with the soft expansion performed in powers of $E/m$, where
$E$ is the soft energy and $m$ the mass of the emitter. Clearly the corresponding
derivation does not apply for particles of vanishing mass. In modern language, 
the problem is that in the massless limit collinear divergences arise, governed
by a different physical scale with respect to the hard scale of the process. In
dimensional regularization, these divergences generate logarithms of that scale,
both at LP and NLP, which cannot be captured by the soft expansion. As an illustration,
consider the cut graph displayed in Fig.~1, contributing to the Drell-Yan
cross section at two loops: near threshold, gluon $k_2$ is always (next-to-) soft,
but gluon $k_1$ is virtual and its momentum components are unconstrained.
When $k_1$ is (next-to-) soft, the contributions of this diagram are correctly 
captured by \eq{neikexp}, but when it is hard and collinear to $p$ it contributes
to LP and NLP logarithms at every order in the soft expansion.
\begin{figure}
\begin{center}
\includegraphics[width=.5\textwidth]{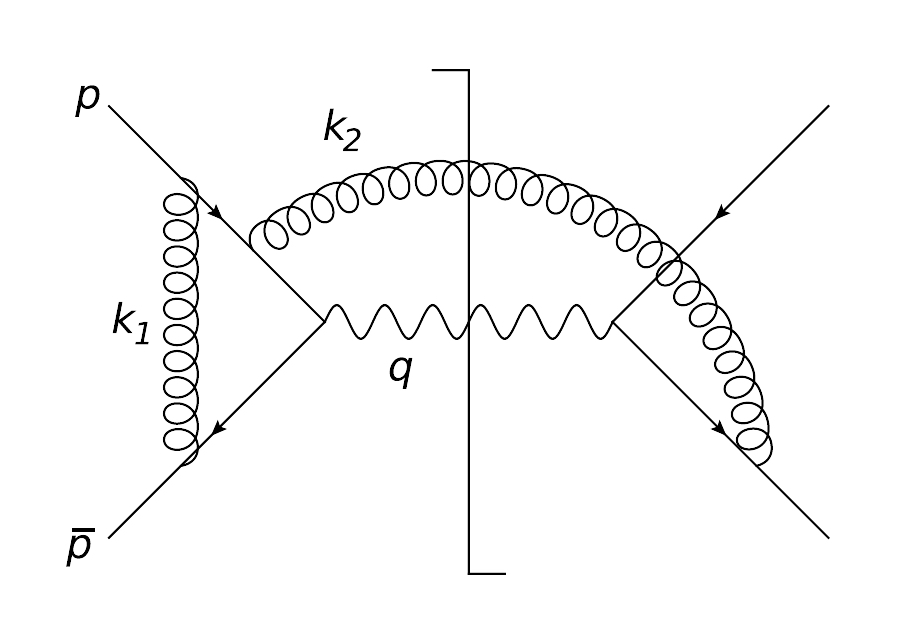}
\caption{A two-loop cut diagram displaying a collinear singularity affecting Low's theorem.}
\end{center}
\label{graph}
\end{figure}

This problem was known since early days, and (in the case of QED) it was solved by 
Del Duca in Ref.~\cite{DelDuca:1990gz}. The solution provides a generalization of
Low's theorem (which we refer to as LBKD theorem) valid for soft energies in the
range $m^2/Q < E < m$, with Q the hard scale of the problem. Clearly, this generalization
applies to the massless limit, and it can be adapted to QCD. With minor modifications, 
the LBKD theorem expresses a hard amplitude with the radiation of an extra soft gluon
in terms of the non-radiative amplitude and of two universal jet functions organising
the collinear enhancements. We write it as 
\beqa
\label{NEfactor2}
  {\cal A}^\mu (p_j, k) & = & \sum_{i = 1}^2 \Bigg\{ \, q_i \left( \frac{(2 p_i - k)^\mu}{2 p_i 
  \cdot k - k^2} + G^{\nu \mu}_i \, \frac{\partial}{\partial p_i^\nu} \right) \\
  & & \hspace{1cm} + \, G^{\nu \mu}_i \left[ \frac{J_\nu (p_i, k, n_i)}{J(p_i, n_i)} - q_i \, 
  \frac{\partial}{\partial p_i^\nu} \Big( \ln J(p_i, n_i) \Big) \right]
  \Bigg\} \, {\cal A} (p_i; p_j) \, . \nonumber
\eeqa
One easily recognizes in the first line of \eq{NEfactor2} the eikonal factor, supplemented
with NE corrections (one could, of course, expand the first term in powers of $k$, or set
$k^2 = 0$ for on-shell real radiation). The second term on the first line corresponds to Low's 
theorem in the absence of collinear enhancements. Indeed, $G_i^{\mu \nu}$ is a tensor 
associated with the hard leg carrying momentum $p_i$ and defined by
\beq
G_i^{\mu \nu} \, = \, \eta^{\mu \nu} - \, K_i^{\mu \nu} \, ; \qquad 
K_i^{\mu \nu} = \frac{(2 p_i - k)^\nu}{2 p_i \cdot k - k^2} \, k^\mu \, .
\label{Gmunu}
\eeq
$G_i^{\mu \nu}$ satisfies $p_i^\mu G_{i, \mu \nu} \sim k_\nu$, thus suppressing
collinear configurations. The second line of \eq{NEfactor2} contains the collinear 
enhancements, collected in the two jet functions $J$ and $J_\mu$. The non-radiative 
jet function $J(p_i, n_i)$ is responsible for collinear divergences along the direction
$p_i$ in the factorised non-radiative amplitude~\cite{Dixon:2008gr}. It is defined by 
the gauge-invariant matrix element
\beq
J (p, n) u(p) \, = \, \langle 0 | \Phi_n \left(0, \infty \right) \psi(0) | p \rangle \, ,
\label{jet}
\eeq
with $n^\mu$ a reference direction for the Wilson line $\Phi_n$, and $\psi$ the 
quark field. The radiative jet function $J_\mu (p_i, n_i, k)$, on the other hand, appears
for the first time in this context. It is defined by the gauge-invariant matrix element
\beq
  J_\mu \left( p, n, k \right) u(p) \, = \, 
  \int d^d y \,\, {\rm e}^{ - {\rm i} (p - k) \cdot y} \, \left\langle 0 \left|  T \Big[ \,
   \, \Phi_{n} (y, \infty) \, \psi (y) \, j_\mu (0) \Big]\, \right| p \right\rangle \, ,
\label{Jmudef}
\eeq
where $j_\mu$ is the quark current. The non-radiative jet function is responsible
for the non-factorised, collinearly enhanced next-to-soft emission from the hard 
parton carrying momentum $p$. It can easily be evaluated at tree-level, with the 
result
\beq
  J^{\nu(0)} \left(p, n, k \right) \, = \, - \, \frac{p^\nu}{p \cdot k} + \frac{k^\nu}{2 p \cdot k}
 - \frac{{\rm i} \, k_{\alpha} \Sigma^{\alpha \mu}}{p \cdot k} \, ,
\label{Jnu0}
\eeq
where $\Sigma^{\alpha \mu}$ are the spin one-half generators of the Lorentz 
group.

Eq.~(\ref{NEfactor2}) is still not fully satisfactory, since it contains residual dependence
on the `factorisation vectors' $n_i$, which would need to be subtracted or reabsorbed
into a suitably defined matching coefficient. For the specific case of processes which
are electroweak at tree level (and thus include only two hard colored partons), there 
is however a simpler solution. Ordinarily, one would take the $n_i$'s such that $n_i^2 
\neq 0$, in order to avoid spurious collinear divergences in the $n_i$ direction. In the 
case at hand, however, one may observe that the factor in square brackets in the second 
line of \eq{NEfactor2} is renormalization group invariant: indeed, the UV divergences
of $J_\mu$ cancel those of $J$ in the first term, while the second is UV finite. One 
may therefore evaluate the square bracket in terms of bare quantities, and at this point
it becomes clearly advantageous to pick $n_i^2 = 0$, since with that choice $J(p_i, n_i) 
= 1$. Finally, one can make a natural and physical choice for the two factorisation 
vectors: since the only two physical vectors in the problem are $p_1$ and $p_2$,
one may choose\footnote{In the multi-parton case, one would have a jet for each 
external hard particle with momentum $p_i$. The natural choice then would be to
pick $n_i$ as the direction anti-collinear to  $p_i$.} $n_1 = p_2$ and $n_2 = p_1$. 
With massless reference vectors, \eq{NEfactor2} takes the considerably simpler form
\beq
  {\cal A}^\mu (p_j, k) \, = \, \sum_{i = 1}^2 \left(q_i \, \frac{(2 p_i - k)^\mu}{2 p_i 
  \cdot k - k^2} + q_i \, G^{\nu\mu}_i \frac{\partial}{\partial p_i^\nu} + G^{\nu\mu}_i 
  J_\nu (p_i, k) \right) {\cal A} (p_i; p_j) \, ,
\label{NEfactor3}
\eeq
which can be directly used to compare with perturbative data. The radiative jet 
function for quarks was computed at one loop (for the $C_F$ color structure) 
in~\cite{Bonocore:2015esa}. As a non-trivial test of Eqs.~(\ref{NEfactor2}) and 
(\ref{NEfactor3}), the $C_F^2$ terms of the two-loop real-virtual contribution to 
the Drell-Yan K-factor were reproduced, allowing also for a detailed mapping to 
the method-of-regions calculation of~\cite{Bonocore:2014wua}. The calculation is
reviewed in~\cite{Bonocore:2015anl}.

\section{Perspective}
\label{persp}

With \eq{NEfactor2}, all the conceptual ingredients required to set up a resummation
formalism for NLP threshold logarithms, at least for processes with electroweak 
final states, are in place. The key information embodied in \eq{neikexp} and in 
\eq{NEfactor2} is that, at the amplitude level, the contributions generating NLP 
logarithms are universal in nature, and factorise from the radiationless process, 
at least in the sense that they can be computed by acting on the non-radiative amplitude
either multiplicatively or by means of a differential operator. Much technical work,
however, remains to be done: first the full non-abelian generalisation of \eq{NEfactor2}
must be worked out, together with the renormalisation properties of the radiative jet
function in generic color representations. Then one must move from amplitudes 
to cross sections, which will require the phase space analysis of Ref.~\cite{Laenen:2010uz}.
At that stage, the ordinary techniques of factorisation and evolution can be employed
to construct a complete resummation formula. Finally, the formalism will have to be 
extended to colored final states, which will involve the treatment of hard next-to-collinear
contributions to final state jets. Along the way, a range of phenomenological applications
to interesting collider processes will become available, as has been the case for 
leading-power threshold resummation.

\section*{Acknowledgments}

\noindent Work supported by the Research Executive Agency (REA) of the European 
Union under the Grant Agreements number PITN-GA-2010-264564 (LHCPhenoNet) 
and PITN-GA-2012-316704 (HIGGSTOOLS); by MIUR (Italy), under contract 
2010YJ2NYW$\_$006; by the University of Torino and the Compagnia di San 
Paolo under contract ORTO11TPXK; by the Netherlands Foundation for 
Fundamental Research of Matter (FOM) programme 104, ``Theoretical Particle 
Physics in the Era of the LHC''; by the Dutch National Organization for Scientific 
Research (NWO); by the UK Science and Technology Facilities Council (STFC);
by the Higgs Centre for Theoretical Physics at the University of Edinburgh.

\end{document}